\documentclass[prl,twocolumn,showpacs]{revtex4}

\usepackage{graphicx} %
\usepackage{bm}       
\usepackage{dcolumn}  
\usepackage{amsmath}

\begin{document}

\preprint{NIMS-MA-04-04}

\title{\textit{Ab initio} calculations of $H_{c2}$ for Nb, NbSe$_{2}$, and 
MgB$_{2}$}

\author{Masao Arai}
\email{arai.masao@nims.go.jp}
\affiliation{
 Computational Materials Science Center, National Institute for Materials Science,
  Tsukuba, 305-0044, Japan }
\author{Takafumi Kita}
\affiliation{Division of Physics, Hokkaido University, Sapporo 060-0810, Japan}

\date{today}

\begin{abstract}
  We report on quantitative calculations of the upper critical field $H_{c2}$ for clean type-II
  superconductors Nb, NbSe$_{2}$, and MgB$_{2}$ using Fermi surfaces from {\em ab initio} 
  electronic structure calculations. The results for Nb and NbSe$_2$ excellently reproduce
  both temperature and directional dependences of measured $H_{c2}$ curves without any adjustable 
  parameters, including marked upward curvature of NbSe$_{2}$ near $T_{c}$. 
  As for MgB$_2$, a good fit is obtained for a $\pi$/$\sigma$ gap ratio 
  of $\sim\! 0.3$, which is close to the value from a first-principles 
  strong-coupling theory [H.\ J.\ Choi \textit{et al}.\ Nature,\ \textbf{418} 758 (2002)]. Our
  results indicate essential importance of Fermi surface anisotropy for describing $H_{c2}$.
\end{abstract}


\maketitle

Quantitative descriptions of nature form an integral part of physics. In this context, the 
density-functional theory for normal-state electronic structures has substantially enhanced 
our understanding on materials and also made materials design possible \cite{LDA}. The purpose of 
the present paper is to extend those calculations in the normal state to type-II superconductors 
in magnetic fields to perform systematic calculations of $H_{c2}$.

The upper critical field $H_{c2}$ is one of the most fundamental quantities in type-II
superconductors. From the early stage of research on type-II superconductors \cite{hohenberg67}, 
it has been recognized that Fermi surface anisotropy has significant effects on $H_{c2}$.
However, most of the calculations performed so far on $H_{c2}$ have used simplified model
Fermi surfaces and/or phenomenological fitting parameters. Due to this lack of 
{\em ab}-{\em initio}-type calculations, our understanding on $H_{c2}$ remains at a rather 
primitive level. The only exception is the work by Butler on bcc Nb \cite{butler80},
where he obtained an excellent agreement with experiments \cite{williamson70,kerchner80}
by using Fermi surfaces and electron-phonon interactions from 
his {\em ab initio} calculations.
Extending this calculation to other materials is expected to improve our understanding on $H_{c2}$
considerably.

Recently we have derived an $H_{c2}$ equation applicable to low-symmetry crystals, including
Fermi surface anisotropy, gap anisotropy, impurity scatterings, and strong electron-phonon
interactions \cite{kita04}. We here concentrate on the role of Fermi surface anisotropy and 
carry out clean-limit weak-coupling calculations of $H_{c2}$ for Nb, NbSe$_{2}$, and MgB$_{2}$. 
Sauerzopf {\em et al}.\ \cite{sauerzopf87} performed a careful experiment on $H_{c2}$ of 
high-purity Nb to report some discrepancies from preceding experiments 
\cite{williamson70,kerchner80} and the theory by Butler \cite{butler80}. 
Also, $H_{c2}$ curves of NbSe$_{2}$ \cite{toyota76,dalrymple84,sanchez95} and MgB$_{2}$ 
\cite{budko01,zehetmayer02,angst02,welp03} remains essentially unexplained 
quantitatively. 
We here treat Fermi surface anisotropy without any adjustable parameters by using energy 
band data from first-principles electronic structure calculations. 

We have determined $H_{c2}$ by requiring that the smallest
eigenvalue of the following matrix be zero \cite{kita04}:
\begin{equation}
  {\cal A}_{NN'} \!\equiv\! \delta_{NN'} \ln \frac{T}{T_{c}}
 \! +\! 2 \pi T  \sum_{n=0}^{\infty}\left[ \frac{\delta_{NN'}}{\varepsilon_{n}} 
  \!-\! \langle {\phi^{2} \cal K}_{NN'} \rangle \right] . 
  \label{Hc2Equation}
\end{equation}
Here $N$ denotes the Landau level in expanding the pair potential
and $\varepsilon_n \!\equiv\! (2n\!+\! 1)\pi T$ is the Matsubara frequency 
($k_{\rm B}\!=\!\hbar\!=\! 1$).
The function $\phi(\bf{k}_{\rm F})$ specifies gap anisotropy,
which is normalized in such a way that
the Fermi-surface average $\langle\phi^{2}\rangle$ is equal to unity.
The matrix ${\cal K}\!=\!{\cal K}(\varepsilon_{n},{\bf k}_{\rm F},H)$ is given explicitly 
in Ref.~\onlinecite{kita04}.
We need to evaluate the average $\langle {\phi^{2} \cal K} \rangle$ 
in Eq.\ (\ref{Hc2Equation}) appropriately.

We have obtained Fermi surfaces and velocities from electronic structure calculations 
within the local density approximation (LDA) \cite{LDA},
using the WIEN2k package \cite{WIEN2k} 
which is based on the full-potential linear augmented 
plane wave method. 
The self-consistent calculations have been performed by using finite ${\bf k}$ points.
Then we have fitted the energy bands by linear combinations 
of the star functions \cite{koelling86, pickett88,arai03} to construct Fermi surfaces
on finer mesh points. 
Integrations over Fermi surfaces have been performed by the linear 
tetrahedron method \cite{jepsen71}.
We have checked the convergence by increasing ${\bf k}$ points.

We shall focus on the temperature and directional dependences 
of $H_{c2}$ and use
the reduced quantities $t\!\equiv\! T/T_{c}$ and
$h^{*}(t) \!\equiv\! \frac{{H}_{c2}(t)}{-d{H}_{c2}(t)/dt |_{t=1} }$.
When it becomes relevant, the absolute value of $H_{c2}$ 
is fixed by using an experimental value at a particular temperature and field direction.
This procedure is similar in the normal state to incorporating many-body mass enhancement
by using a single point of de Haas-van Alphen data.
We have fixed $\phi({\bf k}_{\rm F}) \!=\! 1$ 
for Nb and NbSe$_2$, since gap anisotropy may not be substantial in these materials.  
The spin-orbit interaction have been included self-consistently for NbSe$_2$.

\begin{figure}[tb]
  \includegraphics[width=7.5cm]{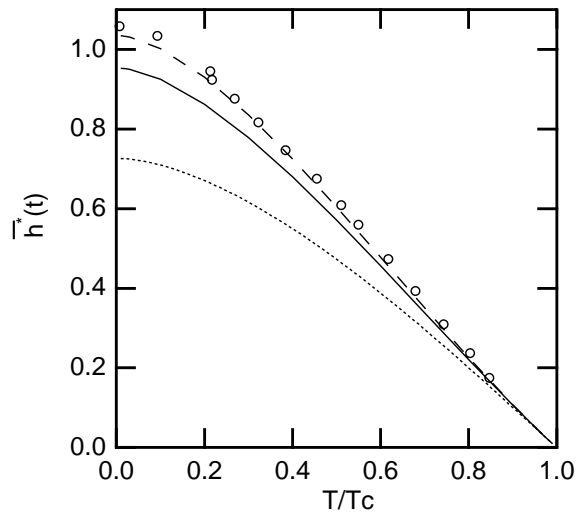}
  \caption{Normalized angle averaged $H_{c2}$ for Nb. 
  Open circles are the experiment of Ref.~\onlinecite{sauerzopf87}. 
  The solid line is obtained by using the Fermi surface from
  the LDA calculation, whereas
  the dotted line is the Helfand-Werthamer curve for the spherical Fermi surface.
  The dashed line incorporates the difference in mass renormalization 
  among different Fermi surface sheets.}
  \label{fig:nb-hc2}
\end{figure}

\textit{bcc N\lowercase{b}} --- 
We first consider temperature dependence of $H_{c2}$ in Nb.
Figure \ref{fig:nb-hc2} shows theoretical curves for the angle averaged critical field  
$\bar{h}^{*}(t)$ in comparison with the experiment
by Sauerzopf {\em et al}.\  (circles) \cite{sauerzopf87}.
The solid line with $\bar{h}^{*}(0)\!=\!0.96$ has been obtained as above
by using the Fermi surface from the LDA calculation, 
whereas the dotted line with $\bar{h}^{*}(0)\!=\!0.73$ is the 
Helfand-Werthamer result \cite{helfand66} for the spherical Fermi surface.
Considering that there are no adjustable parameters,
our result (solid line) shows a fairly good agreement with the experiment,
improving on the Helfand-Werthamer curve substantially.
However, the value $\bar{h}^{*}(0)\!=\!0.96$ still falls well below the
experimental value $\bar{h}^{*}(0)\!=\!1.06$.

We attribute this discrepancy to the difference in
many-body mass enhancement among different Fermi-surface sheets.
According to Butler's calculation \cite{butler80}, 
strong but ${\bf k}$-independent electron-phonon interactions 
cannot cure the discrepancy, 
since they merely bring an overall enhancement of $H_{c2}$ but
change temperature and directional dependences of $\bar{h}^{*}(t)$ by
less than 2\%.
On the other hand, Crabtree \textit{et al}.\ \cite{crabtree87} found 
from their analysis on de-Haas van Alphen experiments
that the electron-phonon renormalization factor $\lambda_{n}({\bf k}_{\rm F})$ 
differs substantially
by more than 30\% among the three Fermi-surface sheets ($n\!=\!1,2,3$) of Nb,
although it does not vary appreciably within each sheet.
Hence we have performed another calculation
of $H_{c2}$ using the scaled Fermi velocity 
$\tilde{\bf v}_n({\bf k}_{\rm F}) \!=\! {\bf v}_n({\bf k}_{\rm F})/(1 \!+\! \lambda_n)$,
where ${\bf v}_n({\bf k}_{\rm F})$ is the bare Fermi velocity from the LDA calculation
and $\lambda_n$ is a sheet-dependent scaling factor taken from Ref.~\onlinecite{crabtree87}.
This yields the dashed line in Fig.~1 with $\bar{h}^{*}(0)\!=\! 1.03$,
showing a considerable improvement on the bare result.
This fact indicates the importance of band and momentum dependences of 
the electron-phonon couplings.
It is an interesting problem in the future to see whether the
effects are reproduced naturally through
a first-principles strong-coupling calculation of $H_{c2}$.


We next discuss angular dependence of $H_{c2}$ in Nb.
This dependence for cubic materials has been studied
by expanding the relative anisotropy $H_{c2}/\bar{H}_{c2}$ as \cite{seidl78,sauerzopf87}
\begin{equation}
  {H_{c2}(\Omega, t)}/{\bar{H}_{c2} (t)} = 1 + 
  \sum_{l = 4, 6, 8,\cdots} a_l(t) P_l(\Omega) \, .
\end{equation}
Here $\Omega$ denotes direction of the field and
$P_l$'s are cubic harmonics \cite{large47} invariant under cubic symmetry operations and
constructed as linear combinations of the spherical harmonics.
The coefficients $a_l(t)$ specify the $H_{c2}$ anisotropy at temperature $t$.
After the work of Butler \cite{butler80}, Sauerzopf {\em et al}.\ \cite{sauerzopf87}
carried out a very careful and detailed experiment on 
$a_l(t)$.
Hence it is very interesting to see whether $a_l$'s can also be reproduced by
{\em ab initio} calculations.
We have evaluated $a_l$'s by least square fittings 
for $H_{c2}$ computed over 50 different directions.

Figure 2 compares calculated $a_l(t)$ ($l \!=\! 4, 6, 8, 10$)
with the experiment by Sauerzopf {\em et al}.\ \cite{sauerzopf87}.
The agreements are good, especially for the dominant $a_4(t)$. 
For example, the change in the curvature of $a_4(t)$ around
$t \!\approx\! 0.6$ is excellently reproduced.
This is not the case for the second-largest contribution $a_6(t)$, however, and 
the calculated curve reproduces only 60\% 
of the experimental value at $t \!=\! 0$. 
An improvement is observed by introducing the Fermi-surface-dependent
renormalization factor $\lambda_{n}$ in the calculation (dashed line),
but there still remains an appreciable deviation.
This discrepancy may be attributed to the small gap anisotropy 
which may be present in Nb.
Indeed, Seidl \textit{et al}.\ \cite{seidl78} reported 
that the gap anisotropy has a dominant effect on the $a_6$ term.
However, their analysis may not be reliable quantitatively, 
especially near $t\!=\! 0$, 
since they applied a theory by Teichler \cite{teichler75}
which incorporate only the lowest-order corrections
from the gap and Fermi-surface anisotropies.
The discrepancy may be removed completely by 
a first-principles strong-coupling calculation of $H_{c2}$
where the gap anisotropy is naturally included.

\begin{figure}[tb]
  \includegraphics[width=8.0cm]{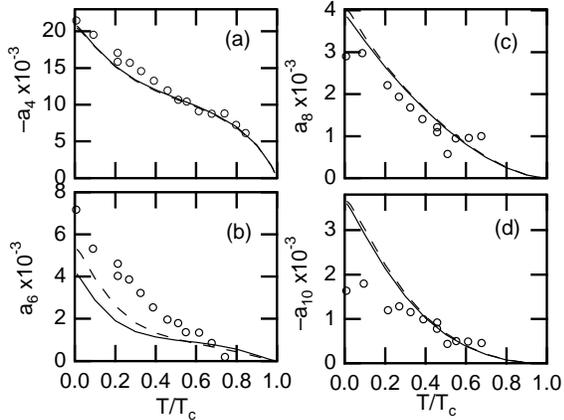}
  \caption{Anisotropy coefficients 
  $a_l$ of Nb as a function of temperature
  for $l \!=\! 4, 6, 8, 10$.
  Solid lines are obtained by a bare LDA calculation, whereas 
  dashed lines denote results using the renormalized Fermi velocity 
  $\tilde{\bf v}_n({\bf k}_{\rm F}) \!=\! {\bf v}_n({\bf k}_{\rm F})/(1 \!+\! \lambda_n)$.
  Open circles are experiments by 
  Sauerzopf {\em et al}.\ \cite{sauerzopf87}. }
  \label{fig:nb-an}
\end{figure}

\textit{N\lowercase{b}S\lowercase{e}$_2$} ---
The hexagonal transition metal dichalcogenide 2H-NbSe$_2$ has been studied extensively as
a prototype of anisotropic layered superconductors. 
Nb atoms form triangular lattices in the $ab$ plane,
which stack along the $c$ axis with two Se layers between them. 
At low temperatures preceding superconductivity, 
NbSe$_2$ undergoes an incommensurate charge-density-wave (CDW) transition. 
Several groups \cite{toyota76,dalrymple84,sanchez95} studied 
$H_{c2}$ of NbSe$_2$ and Nb$_{1-x}$Ta$_x$Se$_2$, where 
$H_{c2\parallel}$ for the magnetic field along the $ab$ plane was observed
by more than 3 times larger than $H_{c2\perp}$ for the field along 
the $c$ axis at $T \! =\! 0$. 
Another characteristic feature is the marked positive curvature 
of $H_{c2\parallel}$ near $T_c$. 
However, no quantitative calculations using detailed Fermi surfaces have been performed yet.

We have evaluated $H_{c2}$ using the Fermi surface from the LDA calculation, ignoring
any effects of the CDW instability.
Calculated energy bands and Fermi surface agree well 
with a previous one \cite{corcoran94}. 
The Fermi surface can be classified into two groups.
The first one is cylindrical sheets, and 
the other is a small hole pocket around $\Gamma$ point which looks like a flattened spheroid.

\begin{figure}[tb]
  \includegraphics[width=7.5cm]{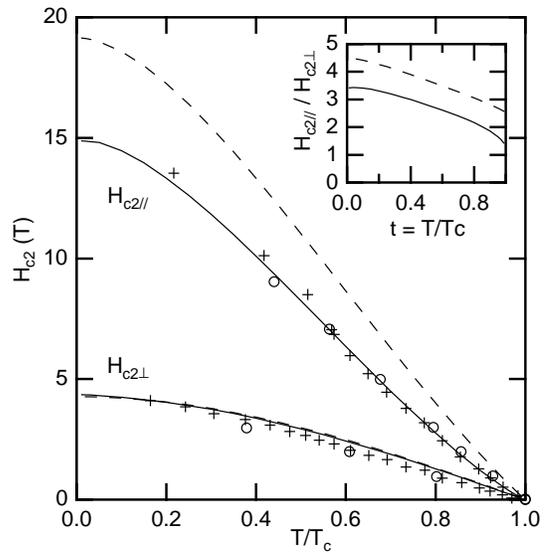}
  \caption{
  $H_{c2\parallel}$ and $H_{c2\perp}$ of NbSe$_2$ as a function of temperature.
  Crosses and open circles are experiments of Ref.~\onlinecite{toyota76} and 
  Ref.~\onlinecite{sanchez95}, respectively.
  The solid lines are from the LDA calculation. 
  If we ignore the Fermi surface around $\Gamma$ point, the dashed lines result.
  The inset plots calculated anisotropy ratio
  $H_{c2 \parallel}/H_{c2 \perp}$.
   }
  
  \label{fig:nbse2-hc2}
\end{figure}

Figure \ref{fig:nbse2-hc2} shows calculated $H_{c2\parallel}$ and $H_{c2\perp}$ curves
in comparison with a couple of experiments \cite{toyota76,sanchez95}.
Temperature dependence of the anisotropy parameter 
$\gamma \!\equiv\! H_{c2\parallel}/H_{c2\perp}$ is also shown in the inset.
We observe good agreements between the theory and experiments for both field directions.
Especially, the positive curvature of $H_{c2\parallel}$ has been reproduced excellently.
Although this positive curvature has been observed commonly in anisotropic superconductors 
\cite{woollam74},
its origin has not been identified clearly so far.
Thus, we have shown explicitly that Fermi surface anisotropy adds quite a variety to 
$H_{c2}$ curves. 
The reduced critical field $h^{*}(t)$ is also very anisotropic,
reaching $h^{*}_{\parallel}(0)\!=\!1.73$ and $h^{*}_{\perp}(0)\!=\!0.68$
in our calculation.
If we ignore the small Fermi surface sheet around $\Gamma$ point, 
the agreement becomes worse as shown by the dashed line. 
Thus, the small Fermi surface sheet cannot be neglected for
the quantitative understanding of $H_{c2}$.

\textit{M\lowercase{g}B$_2$} --- 
Superconductivity in MgB$_{2}$ found by Nagamatsu {\em et al}.\
\cite{nagamatsu01} has attracted much attention. 
Besides its high transition temperature $\sim\! 40$K, 
it has a unique feature that the energy gap $\Delta$ differs
substantially in magnitude between $\sigma$- and $\pi$-bands,
as predicted by theories \cite{liu01,choi02} and 
confirmed by experiments \cite{wang01,bouquet01,szabo01}.
A first-principles calculation by Choi \textit{et al}.\ \cite{choi02} 
has yielded $\Delta_\sigma\!\sim\! 6.8$meV and $\Delta_\pi\!\sim\!1.8$meV 
on the average as $T\!\rightarrow\! 0$.

Experiments on $H_{c2}$ of uniaxial MgB$_2$ single crystals have
been reported by several groups \cite{budko01,zehetmayer02,angst02,welp03}.  
Miranovi\'c \textit{et al}.\ \cite{miranovic03} performed a calculation
of $H_{c2}$ adopting a model Fermi surface with two spheroids, which yielded
a qualitative agreement with observed $H_{c2}$.
We have carried out a more detailed calculation using the Fermi surface from the LDA
calculation. As seen above, detailed Fermi-surface structures are indispensable for 
the quantitative understanding of $H_{c2}$.
As for the gap anisotropy, we have incorporated the ratio 
$\alpha\!\equiv\! \Delta_{\pi}/\Delta_{\sigma}$ at $T\!=\!0$
as a single parameter in the calculation.

\begin{figure}[tb]
  \includegraphics[width=7.5cm]{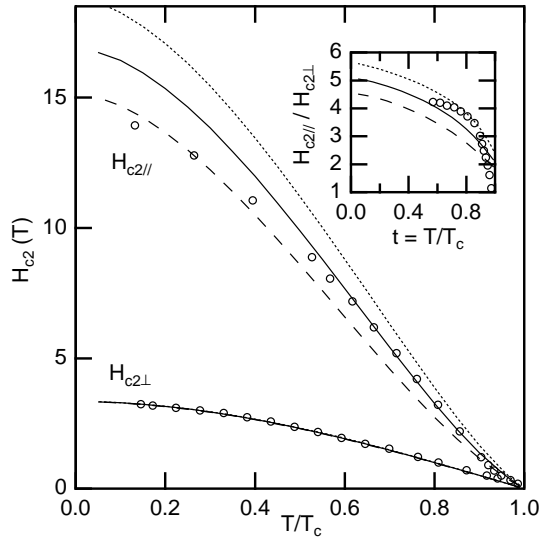}
  \caption{
  $H_{c2}$ of MgB$_2$ as a function of temperature.
  Open circles are experiments of Ref.~\onlinecite{zehetmayer02}.
  Calculated $H_{c2}$ curves for $\Delta_{\pi}/\Delta_{\sigma} = 0.25$, $0.3$, and $0.35$
  are shown by dotted, solid and dashed lines, respectively.
  The inset shows a comparison of the anisotropy ratio $H_{c2 \parallel}/H_{c2 \perp}$.
  }
  \label{fig:mgb2-hc2}
\end{figure}

Figure \ref{fig:mgb2-hc2} plots calculated $H_{c2\parallel}$  
and $H_{c2\perp}$ for MgB$_2$ using
three different values of $\alpha$.
When we assume an isotropic gap of $\alpha\!=\! 1$, 
we could not describe the experimental anisotropy $H_{c2\parallel}/H_{c2\perp}$. 
The anisotropy of $0.6 \!\lesssim\! t \!\lesssim \! 1.0$ 
has been reproduced most completely for the choice $\alpha\!\sim\! 0.3$. 
This value is close to the one obtained by Choi {\em et al}.\ \cite{choi02} 
and consistent with the model calculation by Miranovi\'c \textit{et al}.\ \cite{miranovic03}.
Thus, the present calculation provides an additional support for the
existence of two energy gaps via a detailed calculation of $H_{c2}$.
The reduced critical field $h^{*}(t)$ is very anisotropic as
$h^{*}_{\parallel}(0)\!=\!1.64$ and $h^{*}_{\perp}(0)\!=\!0.67$
for $\alpha\!=\!0.3$.
However, calculated $H_{c2}$ for $\alpha = 0.3$ shows 
larger anisotropy at lower temperatures than the experimental one.  
This discrepancy may be removed completely
by incorporating effects neglected here,
such as strong electron-phonon interactions, Pauli paramagnetism,
etc.

In summary, we have evaluated $H_{c2}$ of three classic type-II superconductors using
Fermi surfaces from first-principles electronic structure calculations.
For Nb and NbSe$_{2}$ whose superconducting gap anisotropy may not be significant,
calculated $H_{c2}$ curves satisfactory reproduce experimental
temperature and directional dependences of $H_{c2}$ without any adjustable parameters.
As for MgB$_2$, a good agreement with experiments follows if
we choose the gap ratio $\Delta_{\pi}/\Delta_{\sigma}\!\sim\! 0.3$. 
Thus, the present study has clarified unambiguously the necessity of
incorporating detailed Fermi surface structures in the calculation of $H_{c2}$.
Such calculations will also form a basic starting point to discuss other contributions,
and we may obtain unique information about superconducting gap anisotropy, etc.,  
by comparing calculated and experimental $H_{c2}$ curves. 

\begin{acknowledgments} 
  We would like to thank to K. Kobayashi for valuable discussions.
  This research is supported by a Grant-in-Aid for Scientific Research 
  from Ministry of Education, Culture, Sports, Science, and Technology of Japan.
\end{acknowledgments}
\bibliography{paper03} 

\end{document}